\pdfoutput=1
\documentclass[pre,twocolumn]{revtex4}
\usepackage{times,mathptmx}
\usepackage{graphicx} 
\usepackage[english]{babel}
\usepackage[T1]{fontenc}
\usepackage[colorlinks=False,pdfborder=0 0 0  ]{hyperref}
\usepackage{amssymb}
\usepackage{amsfonts}
\usepackage{amsmath}
\usepackage{bbold}
\usepackage{bm}

\newcommand{\coolimg}[3]{
\begin{figure}[t!]
		\includegraphics[width=1.\columnwidth,keepaspectratio=true]{#1}
		\caption{#2 \label{fig:#3}}
\end{figure}
}

\usepackage{soul}
\newcommand{\GS}[1]{{#1}}
\newcommand{\GSdel}[1]{}

\begin{document}

\title{Conformations, hydrodynamic interactions, and instabilities of sedimenting semiflexible filaments} 

\author{Guglielmo Saggiorato} 
\author{Jens Elgeti}
\author{Roland G. Winkler}
\author{Gerhard Gompper}
\email{g.saggiorato@fz-juelich.de, j.elgeti@fz-juelich.de, r.winkler@fz-juelich.de, g.gompper@fz-juelich.de}
\affiliation{Theoretical Soft Matter and Biophysics, Institute of Complex Systems and Institute for Advanced Simulation, Forschungszentrum J\"{u}lich,
52425 J\"{u}lich, Germany}

\begin{abstract}
The conformations and dynamics of semiflexible filaments subject to a homogeneous external (gravitational)
field, e.g., in a centrifuge, are studied numerically and analytically. The competition between hydrodynamic
drag and bending elasticity generates new  shapes and dynamical features.
We show  that the shape of a  semiflexible filament undergoes instabilities as the external field increases.
We identify two transitions that correspond to the excitation of higher bending modes.
In particular, for strong fields the filament stabilizes in a non-planar shape, resulting in a sideways drift or in helical trajectories.
For two interacting filaments, we find the same transitions, with the important consequence that the new
non-planar shapes have an effective hydrodynamic repulsion, in contrast to the planar shapes which attract  themselves even when their osculating planes are rotated with respect to each other.
For the case of planar filaments, we show analytically and numerically that the relative velocity
is not necessarily due to a different drag of the individual filaments, but to the
hydrodynamic interactions induced by their shape asymmetry.
\end{abstract}
%  \end{@twocolumnfalse}
%  ]

\maketitle

%%%MAIN TEXT%%%%
\section{Introduction}

Semiflexible filaments are fundamental constituents of micro-biological systems, where microtubules and
actin filaments serve as scaffolds for cellular structures and as routes to sustain and guide cellular
transport systems.\cite{howard_mechanics_2001} Microtubules are also the main structural
elements of cilia and sperm flagella, where their relative displacement and deformation due to
motor proteins gives rise to the flagellar beat and hydrodynamic propulsion.\cite{camalet_self-organized_1999,camalet_generic_2000} Microtubules and flagella can be seen as elastic filaments interacting with their own flow field.
The ability to visualize, assemble, and manipulate biological and artificial semiflexible polymers
\cite{arratia_elastic_2006,wiggins_trapping_1998,schroeder_observation_2003,coq_collective_2011}
poses new fundamental questions on the dynamics of filaments when elastic and hydrodynamic forces
compete.

The dragging of stiff rods through a viscous fluid has been studied in detail.\cite{happel_low_1983}
A single rod does not reorient, but falls with its initial orientation. A more complex dynamical
behavior can be expected and is indeed observed for semiflexible filaments when the curvature or stretching elasticity competes with the hydrodynamic interactions.\cite{manghi_hydrodynamic_2006,llopis_sedimentation_2007,schlagberger_orientation_2005}
Single dragged semiflexible filaments bend into a shallow $V$-shape to balance the higher drag at
both ends \cite{cosentino_lagomarsino_hydrodynamic_2005} and their end-to-end vector aligns perpendicularly to the external
field.\cite{schlagberger_orientation_2005}
For strong drag, higher modes  have been found to be excited; this generates
$W$-shapes initially, which then relax back into horseshoe-like $U$-shapes.\cite{cosentino_lagomarsino_hydrodynamic_2005} Here, the dynamics seems to be constrained to
the plane initially defined by the direction of the external field and the filament itself.
However, these investigations address the problem from a deterministic point of view, and
little attention has been paid to the dynamic stability of the resulting shapes.
In all cases, the dragged and deformed semiflexible filament initially defines the settling
plane, but the stability of the filament's planar shape has not been investigated as function
of the external field or the relative position of possible neighboring filaments.

\begin{figure*}
\centering
\includegraphics[width=1.5\columnwidth]{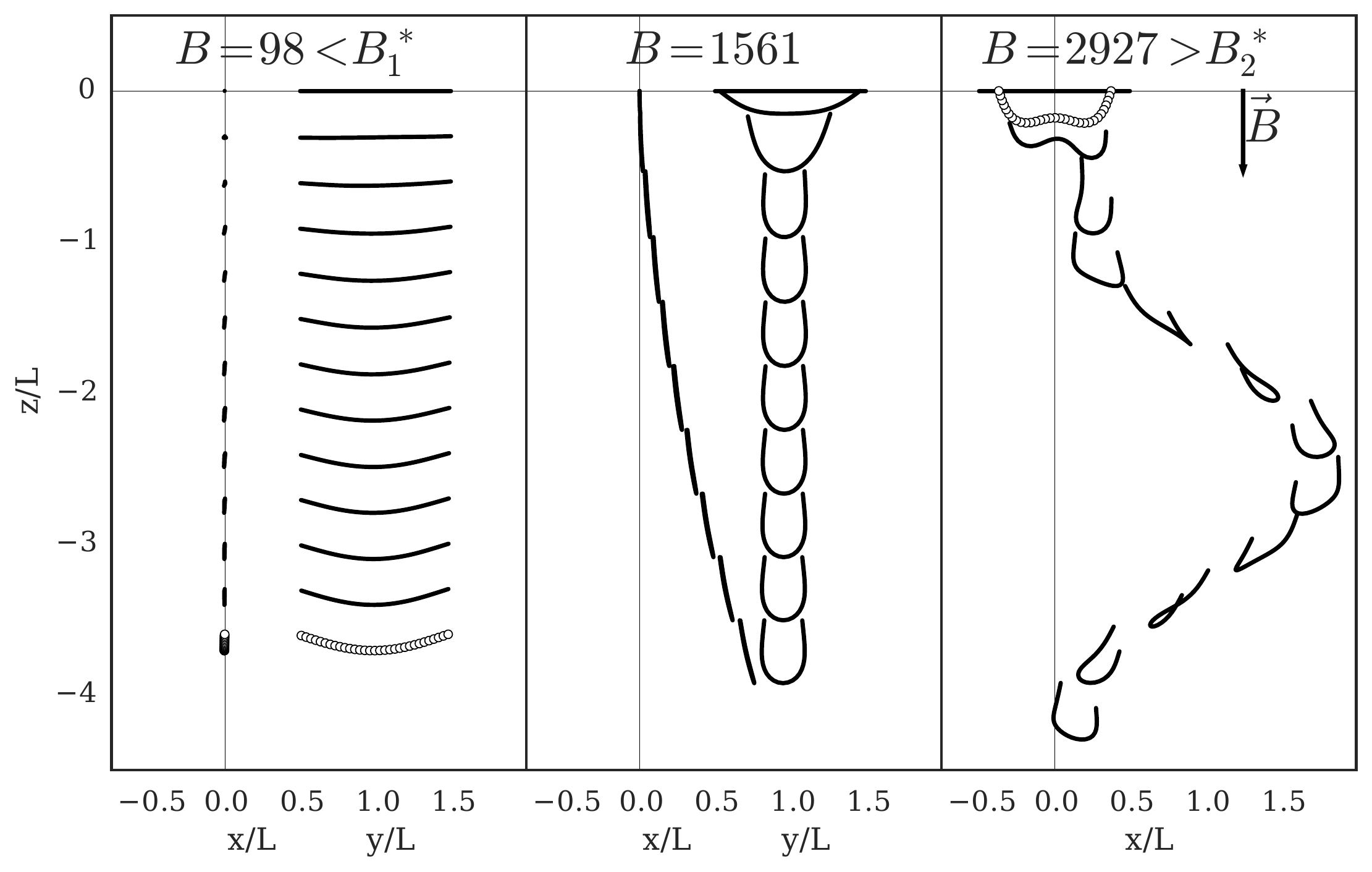}
\caption{\label{fig:geometries}
Snapshots from simulations of single filaments dragged by the  external homogeneous field
$B=mgL^2/\kappa$,
where $L$ is the filament length, $g$ the external field, and $\kappa$ the bending stiffness.
\textbf{Left:} For weak field $(B<B_1^*)$ the filament bends into a $V$-shape (in dots), dominated by
the $\chi_{2,z}$ mode.
\textbf{Center:} As the field strength increases, higher modes with an out-of-plane component are excited,
and the filament drifts sideways.
\textbf{Right:} For even stronger fields $(B>B_2^*)$ further symmetries are spontaneously broken, and
the filament rotates following a helical trajectory. Corresponding movie is shown in the ESI. \GS{The vertical distance between the frames is reduced to enhance the visualization of the filament conformations.}}
\end{figure*}

Here, we focus on the full three-dimensional shape of  one, two, and three semiflexible filaments
sedimenting in a homogeneous external field.
% with a conformation-dependent drag.
We incorporate the hydrodynamics into the equations of motion  for the filament shape via the Oseen tensor,
valid in the limit of zero Reynolds number. As a result of our numerical and analytical analysis,
we find that the deformations confined to a plane become unstable with respect to normal perturbations
at a threshold value $B_1^*$ of the strength $B$ of the external field, which is {\em smaller} than the
threshold $B_2^*$ where initial, transient $W$-shapes become excited, see Fig.~\ref{fig:geometries}.
Thus, with increasing strength of the external field, two instabilities and transitions to new sedimentation
modes are predicted.  The first transition is from a stable planar $U$-shape with little bending to a
stationary horseshoe-like $U$-shape with out-of-plane bending. The second transition at stronger
fields excites a metastable $W$-shape, also with out-of-plane bending, which then ``relaxes'' into
a non-stationary asymmetric $U$-shape.
As result, there exist two families of shapes, where the elastic forces are balanced by a
conformation-dependent drag.

We consider next the interaction between two filaments in an external field.
Indeed, while the dynamics of an isolated filament  is an indispensable
knowledge needed to understand the case of $n>1$ interacting filaments, many situations are characterized by elastic slender objects interacting via the generated flow field:
cilia,\cite{elgeti_emergence_2013,coq_collective_2011} sperm,
\cite{riedel_self-organized_2005,fisher_competition_2010} and \textit{E. Coli} bundles \cite{berg_chemotaxis_1972,reigh_synchronization_2012} are probably the most relevant from a biological point of view.

It is known that the sedimentation behavior of colloids can be quite complex.
The interaction of sedimenting particles has been studied in considerable detail for spherical
colloids.\cite{ekiel-jezewska_stokesian_2008,ekiel-jezewska_class_2014}
Two particles sediment together, but don't follow the direction of the
external field, and  move instead under an angle with respect to it. For more particles,
many different dynamical behaviors can be found,
in particular periodic motions where particles ``dance'' around each other.\cite{ekiel-jezewska_class_2014}

For dragged semiflexible filaments, the dynamical behavior is even more complex.\cite{llopis_sedimentation_2007}
In particular, we show that two filaments  (Fig.~\ref{fig:geometries}) attract each other, repel each other, or spin around the field  depending on the intensity of the external field.

We focus here on the stability of the  sedimentation plane for different field intensities
and on the origin of the relative velocity. In particular, we want to see whether the velocity
difference is due to different shapes or to the broken up-down symmetry.
For even more filaments, the dynamics becomes unsteady at much weaker external field strength than
expected from the two-filaments case.

\section{Model and Methods}

\subsection{Discrete Model}

In the simulation, filaments of length $L=b(N-1)$ are represented by $N$ mass points of mass $m$ connected by harmonic bonds of length $b$, with the potential
\begin{align}
U_b=\frac{k_b}{2} \sum_{i=0}^{N-1} \left(|\textbf{R}_{i}|-b\right)^2,
\end{align}
where $\mathbf{R}_{i}=\mathbf{r}_{i+1}-\mathbf{r}_{i}$ is the vector connecting the consecutive
points, $i \in \{0,\ldots,N\}$, and $k_b$ is the force constant. To account for filament stiffness, we introduce the bending potential
\begin{align}
U=\frac{\kappa}{2b^3} \sum_{i=0}^{N-2}\left(\mathbf{R}_{i+1}-\mathbf{R}_{i}\right)^2,
\end{align}
with the bending rigidity $\kappa$.\cite{myung_self-organized_2014} In addition,  the mass points are exposed to an external gravitational field with the force on a particle
\begin{align}
\mathbf F_i^G = - m g \mathbf{e}_z ,
\end{align}
with the unit vector $\mathbf{e}_z$ along the z axis of the Cartesian coordinate system (cf. Fig.~\ref{fig:geometries}).

Inter- and intrafilament hydrodynamic interactions substantially influence the filament dynamics. Hence, we apply the equation of motion  
\begin{align}
\gamma_0 \left(\dot{\mathbf{r}}_i-\mathbf{u}_i \right) = \mathbf{F}_i^C
\end{align}
for the overdamped dynamics,\cite{doi:86} where, $\mathbf u_i$ is the background flow velocity at the position $\mathbf{r}_i$ of the particle, $\gamma_0 = 3 \pi \eta b$ is its friction coefficient for a fluid of viscosity $\eta$, and $\mathbf{F}_i^C$ is the sum of all conservative forces.\cite{llopis_sedimentation_2007}
We compute  the  background flow velocity explicitly via the Oseen tensor.\cite{doi:86}
Thus, the final equations of motion are
\begin{align}
\dot{\mathbf{r}}_i = \sum_{j=0}^N \mathbf{H}_{ij} \mathbf{F}_j^C , 
\end{align}
with the hydrodynamic tensor
\begin{align}
\mathbf{H}_{ij} &= \mathbf{H}(\mathbf{r}_i-\mathbf{r}_j)\nonumber \\&= (1-\delta_{ij})\left[\frac{3b}{8\gamma_0 r_{ij}}
\left[\mathbf{I}+\mathbf{\hat r}_{ij}\mathbf{\hat r}_{ij}^T\right] \right]+ \frac{1}{\gamma_0} \mathbf{I} 
\end{align}
and $\mathbf{r}_{ij}=\mathbf{r}_i-\mathbf{r}_j$,
$r_{ij}=\mid \mathbf{r}_{ij}\mid$,  and $\mathbf{\hat r}_{ij}=\mathbf{r}_{ij}/r_{ij}$.

If needed, excluded-volume interactions are implemented via a short-range Yukawa potential between points of
different filaments, which implies a minimal effective distance during the simulations.
A low-amplitude white noise is added \GSdel{, which helps} to avoid metastable states. \GS{The noise is not considered to be of thermal origin as it is chosen to be negligible compared to the other hydrodynamic and mechanical forces and barely influences the stationary settling  dynamics for the considered external fields.}\GSdel{, but that is too weak to generate visibly effects the filaments shape.} 

\subsection{Parameters/Methods}
We set the hydrodynamic diameter of a point equal to the bond length, as in the Shish-Kabab model of
Refs.~\cite{doi:86,cosentino_lagomarsino_hydrodynamic_2005,llopis_sedimentation_2007}, thereby fixing the aspect ratio to $b/L$. Lengths are measured in units of the hydrodynamic diameter $b$ and time in units of  $\gamma_0b^3/\kappa$. This choice eliminates the friction coefficient and the bending rigidity from the equations of motion. The force constant for the bonds is set to $k_b b^3/\kappa=1$\GS{, resulting in a maximum extension of $\pm 0.6\%$ of the length over the investigated range of parameters.} 
In these units, the external field strength $m g$ becomes $G=mgb^2/\kappa$. For convenience and an easier comparison with results of Ref.~\cite{llopis_sedimentation_2007}, we characterize the external force by  $B=N^2G$, or $B=mg L^2/\kappa$. For $B\ll 1$, the bending rigidity dominates and the filament is essentially straight. We consider only filaments of length $L=30b$ in the following.
For excluded volume interactions, the minimal effective distance is approximately~$5b$.
\GS{The equations of motion are integrated with an adaptive time-stepping Velocity-Verlet algorithm.~\cite{eastman_openmm_2013,galassi__m_gnu_2013}}

\subsection{Continuum Model} \label{sec:continuum_model}
For an analytical description of the filament dynamics, we adopt a continuum model. The equation of motion of the point $\mathbf{r}^{\nu}(s,t)$ ($-L/2 \le s \le L/2$) along the contour of filament $\nu$ is given by \cite{harnau_dynamic_1996}
\begin{align}
\label{eq:singlefilament}
  \partial_t \mathbf{r}^{\nu}(s,t)= 
 \sum_{\mu} \int_{-L/2}^{L/2} ds'  \mathbf{H}(\mathbf{r}^{\nu}(s)-\mathbf{r}^{\mu}(s')) 
         \mathbf{f}^{\mu}(s') , 
\end{align}
where $\mathbf{f}^{\mu}$ is the external force density  and the index $\mu$ indicates the various filaments. As before, the hydrodynamic tensor $\mathbf{H}(\mathbf{r}^{\nu}(s)-\mathbf{r}^{\mu}(s'))$ comprises the Oseen tensor and the local friction. Explicitly, it reads as
\begin{align} \label{eq:Oseen}
\mathbf{H}(\mathbf{R})=
\frac{1}{8\pi\eta}\frac{\Theta(|\textbf{R}| -b)}{|\textbf{R}|^3}\left[\mathbf{I}\textbf{R}^2 +\textbf{R} \textbf{R}^T \right] + \frac{\delta (\mathbf{R})}{\gamma} \mathbf{I}.   
\end{align}
Here, $\Theta(x)$ is the Heaviside function, $\gamma = 3 \pi \eta$ is the friction per unit length,  and
$\textbf{R}=\textbf{r}^{\nu}(s)-\textbf{r}^{\mu}(s')$.\cite{harnau_dynamic_1996,winkler_models_1994} The force density $\mathbf{f}$ comprises bond, bending, and gravitational forces. In the limit of a rather stiff filament, it can be written as
\begin{align}
\mathbf{f}^{\nu}(s) = l_p k_B T\left( \frac{1}{l_p^2} \frac{\partial^2}{\partial s^2} -  \frac{\partial^4}{\partial s^4} \right) \mathbf{r}^{\nu}(s) + \mathbf{f}^{\nu}_G(s) ,
\end{align} 
with the persistence length $l_p$.\cite{arag:85,winkler_diffusion_2007} In the following, we will neglect the bond term, i.e., the term with the second derivative and focus on bending stiffness only.

The  expansion 
\begin{equation} \label{eq:eigenfunction_expan}
\mathbf{r}^{\nu}(s,t)=\sum_{n=0}^{\infty} \bm{\chi}_n^{\nu}(t)\phi_n(s) 
\end{equation}
in terms of the eigenfunctions $\phi_n$ of the biharmonic operator, i.e.,
\begin{align} \label{eq:eigenvalue}
 l_p k_B T \frac{\partial^4}{\partial s^4}  \phi_n(s) = \frac{\gamma}{\tau_n} \phi_n
\end{align}
with suitable boundary conditions,\cite{arag:85,winkler_diffusion_2007} leads to the equations of motion of the mode amplitudes
\begin{align} \label{eq:modes_nonlinear}
\partial_t \bm{\chi}_{n}^{\nu}= \sum_{\mu}\sum_{l=0}^{\infty}  \mathbf{H}_{nl}^{\nu \mu} \left[ - \frac{\gamma}{\tau_l}\bm{\chi}_l^{\mu}(t)+\mathbf f_{lG}^{\mu}\right] .
\end{align}
The matrix representation of the hydrodynamic tensor is 
\begin{align}
\mathbf{H}_{nl}^{\nu \mu}=&\int_{-L/2}^{L/2} ds ds' \; \phi_n(s)  \mathbf{H}(\mathbf{r}^{\nu}(s),\mathbf{r}^{\mu}(s')) \phi_l(s').
\end{align}
The eigenfunctions $\phi_n(s)$ and relaxation times $\tau_n$ are well known.\cite{harnau_dynamic_1996,wiggins_trapping_1998,winkler_diffusion_2007} For convenience, we summarize them in Appendix~\ref{app:eigenfunctions}. However, Eq.~(\ref{eq:modes_nonlinear}) is nonlinear and thus cannot straightforwardly be solved.
For the current analysis, the second and forth mode are most important; they are responsible for the $V-$ and $W$-shape displayed in Fig.~\ref{fig:geometries}.

To characterize the numerically obtained filament conformations, we calculate the mode amplitudes
\begin{align} \label{eq:amplitudes}
\bm \chi_n^{\nu}(t) = \int_{-L/2}^{L/2} ds \ \mathbf{r}^{\nu}(s,t) \, \phi_n(s) .
\end{align}
The components of the vector $\bm\chi_{n}^{\nu}$ indicate the 
importance of the mode in the Cartesian directions. For example, the mode amplitude $\chi_{2,z}^{\nu}$ measures
how much the filament bends along the $z$ direction into a $V$-like shape.

As for the discrete model, we scale lengths by the filament diameter $b$ and time by $\gamma_0 b^3/k_BTl_p$. The latter is identical to the time scaling of the discrete model, since $\kappa = k_BTl_p$. This implies for the strength of the external force $G=\rho g b^3/k_BTl_p = mg b^2/\kappa$.

\section{Results}

\subsection{Deformation and Dynamics of Single Filament}\label{ssec:singlefilament}

The filament is initially oriented along the $x$ axis of the reference frame. After a certain time, 
the dragged filament reaches a stationary shape and velocity. Examples of conformational sequences for various field strengths are displayed in Fig.~\ref{fig:geometries}. We characterize the shapes via Eq.~(\ref{eq:amplitudes}) in terms of the mode amplitudes. In
Fig.~\ref{fig:noise_single}, the most important stationary amplitudes are presented.
Below a critical field $B_1^*\simeq 1200$, the filament shape is governed by planar modes
(green and black lines), where $\chi_{2,z}$ dominates and, thus, the characteristic $V$-shape appears.

\coolimg{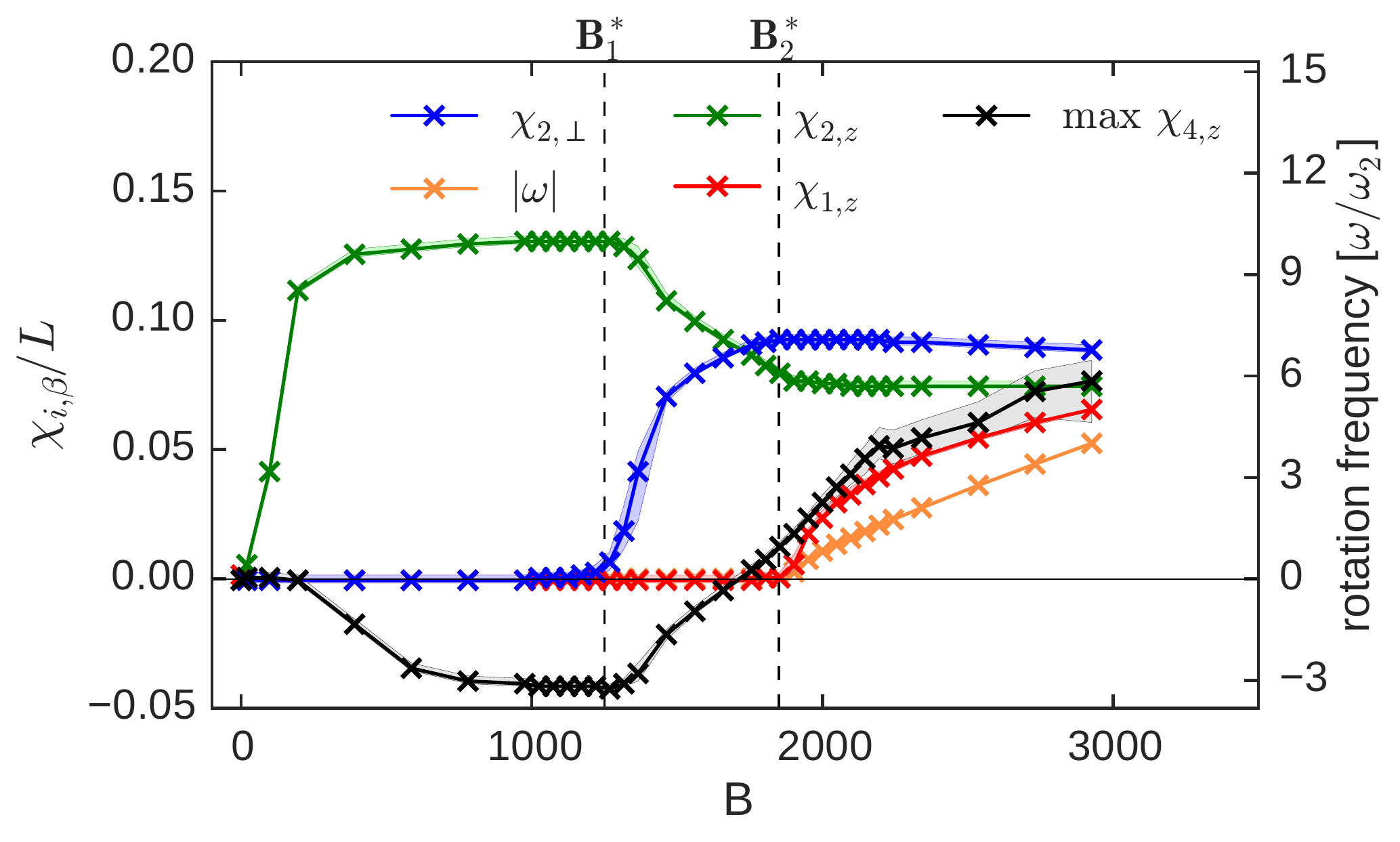}
{Stationary mode amplitudes of a single semiflexible filament as function of the external
field  $B$. The shaded areas indicate  the $66\%$ confidence interval. When $B<B_1^*$,  only planar modes are excited, and the filament stays in the plane defined by its initial orientation and the orientation of the applied field,  here  the $xz$ plane. For $B>B_1^*$, an out-of-plane mode $\chi_{2,\perp}$ is excited.
For $B>B_2^*$, the out-of-plane component $\chi_{2,\perp}$, the bending component
$\chi_{2,z}$ saturates, and the amplitude $\chi_{4,z}$ becomes important
(visualized in Fig.~\ref{fig:geometries}). In black crosses indicate the maximum value of
$\chi_{4,z}$ before it decays. The resulting shape is asymmetric and spirals
around the $z$ axis with frequency $\GS{\mid \omega\mid}/\omega_2$ \GS{(orange line)}, with $\omega_2$ the  frequency of the second mode~(Eq.~(\ref{eq:chi_perp})).}
{noise_single}

In simulations restricted to a two-dimensional plane, or in three-dimensional simulations without noise,\cite{cosentino_lagomarsino_hydrodynamic_2005}
the filament dynamics is localized in the $xz$ plane and filaments bend into
a planar $W$-shape for fields $B>B_2^* \approx 1800$. In contrast,
in our three-dimensional simulations with weak noise, we find that the planer filament conformations
are metastable for $B_1^*< B < B_2^*$, and also modes along the $y$ axis are excited.  
We characterize the out-of-plane filament shape and dynamics by the mode amplitude $\chi_{2,\perp} (t)$, where
\begin{align} \label{eq:chi_perp}
\chi_{2,\perp} (t) = \chi_{2,x}(t) + i \chi_{2,y} (t) = |\chi_{2,\perp}|e^{i \omega t} .
\end{align} 
In the stationary state, an $U$-shaped and deck-chair-like conformation  is assumed with out-of-plane bending (see Fig.~\ref{fig:geometries}). The filament orientation is fixed  and $\chi_{2,\perp}=\chi_{2,y}$ (blue line in Fig.~\ref{fig:noise_single}).
Since its  shape is asymmetric, the filament drifts sideways while settling in the external field.

When $B \gtrsim B_2^*$, the mode $\chi_{4,z}$ becomes important at early times, leading to a temporary $W$-shape (Fig.~\ref{fig:noise_single}). The trajectory for $B \simeq 3000$, displayed in Fig.~\ref{fig:geometries}, shows the initial $W$, which later turns into an asymmetric $U$-shape, in which one arm is longer than the other.
The appearing shape is stable;  however, because of its asymmetry, the mode amplitude $\chi_{1,z}$ is non-zero and the filament rotates around
the $z$ axis with frequency $\omega$, see Fig.~\ref{fig:noise_single} (\GS{orange line}), which we determined via Eq.~(\ref{eq:chi_perp}).

\begin{figure}[t]
\centering
\includegraphics[width=.95\columnwidth]{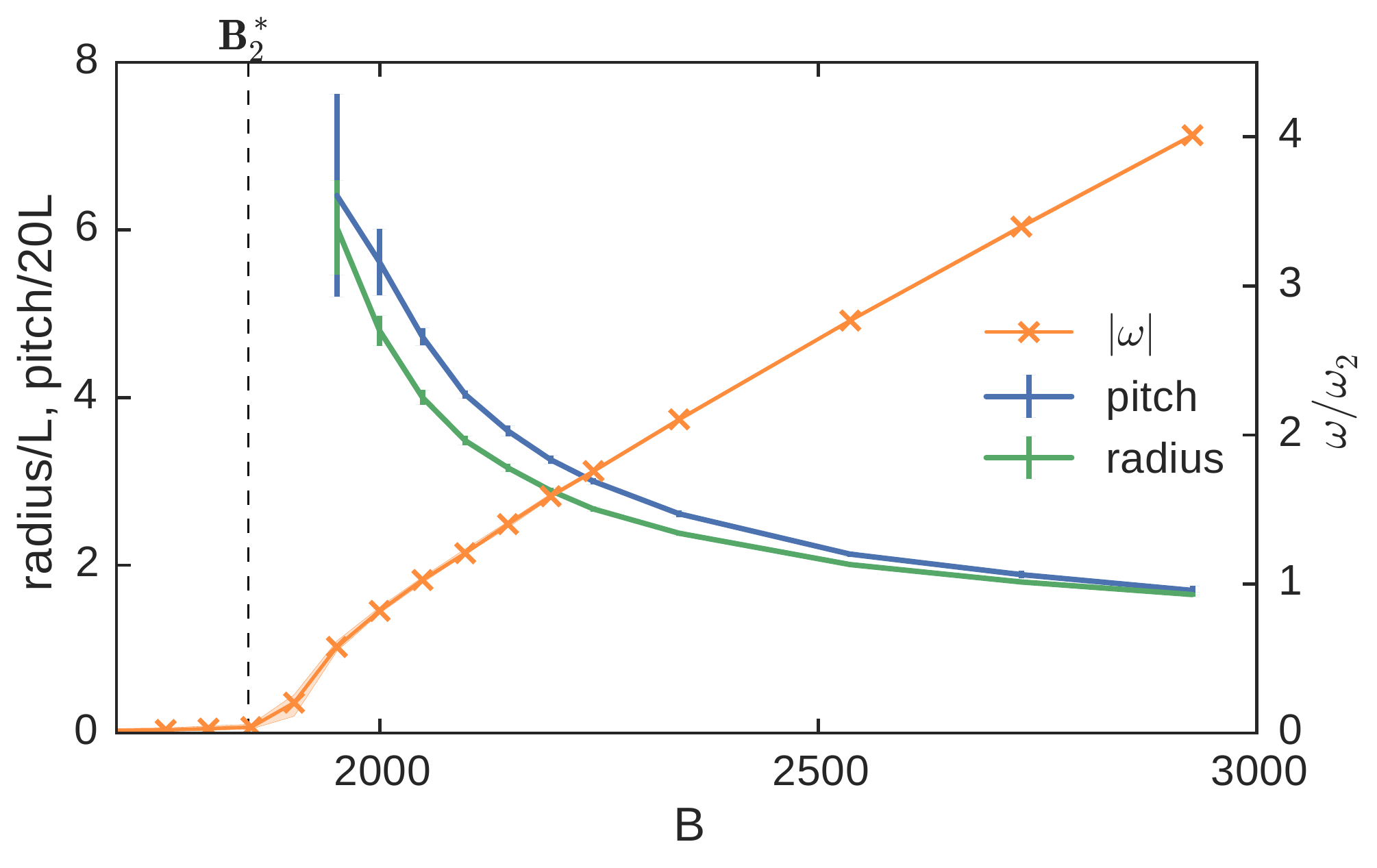}
\caption{\label{fig:addendum} \GS{Pitch, radius, and rotation frequency of the helical trajectories when $B>B_2^*$. The radius and pitch seem to diverge in the proximity of the transition. Bars are standard deviations.}}
\end{figure}

\GS{
In Fig.~\ref{fig:addendum}, we characterize the helical trajectories by the pitch, radius, and rotation frequency ($B>B_2^*$). As the field increases the rotation frequency increases, and the helix becomes more tight because the radius decreases, and the pitch shortens. The ratio between the pitch and the radius defines the helix angle $\alpha\approx4\pi/9$, constant for all $B>B_2^*$. Approaching  the transition point from above, the rotation frequency vanishes, and both radius and pitch diverge because the trajectory straightens.
}

In contrast, in the deterministic dynamics of previous studies,\cite{cosentino_lagomarsino_hydrodynamic_2005} the $W$-shape was
found to decay only into the stable and symmetric planar horseshoe shape.
\subsection{Conformations and Dynamics of Two Interacting Filaments}

\subsubsection{Weak Field - Relative Velocity}

As shown in  Sec.~\ref{ssec:singlefilament}, the stationary shape of a single filament in weak fields $B< B_1^*$ is of $V$-shape, which breaks the bottom-top symmetry. This
is sufficient to generate an effective attraction between sedimenting filaments with the same shape.
To characterize this interaction, we compute the relative velocity $\Delta v$ between the centers of mass of two filaments of equal shape along the sedimentation direction. The filaments remain localized in the $xz$ plane and are  separated by a distance $d$. As shown in Fig.~\ref{fig:fixed_same_shape}, the relative velocities  exhibit a significant dependence on the filament separation. We especially find that $\Delta v \sim d^{-2}$ for distances larger than the filament length.

The distance dependence can be understood by the theoretical model introduced in Sec.~\ref{sec:continuum_model}. For the considered filament shapes,
\begin{align}
\mathbf{r}^1(s,t) = (\chi_{1,x}(t) \phi_1(s), 0, \chi_{2,z} \phi_2(s))^T
\end{align}
and $\mathbf{r}^2(s,t) = \mathbf{r}^1(s,t) + d \mathbf{e}_z$, the general expression for the velocity difference derived in Appendix~\ref{app:relative_velocity} yields
\begin{align} \label{eq:velocity}
\Delta v_{cm} \sim \chi_{2,z}^2\frac{L^2}{d^2} 
\end{align}
in the limit $d \to \infty$. Evidently, the filaments attract each other due to the top-bottom
asymmetry of their shapes. In the simulations, the filament shapes are determined initially by imposing the amplitude $\chi_{2,z}$, which is then kept fixed.
The simulation results of Fig.~\ref{fig:fixed_same_shape} are in agreement with our
theoretical prediction down to roughly the filament length.
The $d^{-2}$ power law is indeed a universal scaling, unaffected by the filaments shape and
external field as evident from the theoretical considerations in Appendix~\ref{app:relative_velocity}. The dependence of $\Delta v$ on  $\chi_{2,z}$ (Eq.~(\ref{eq:velocity})) is also verified for very small bending.
\coolimg{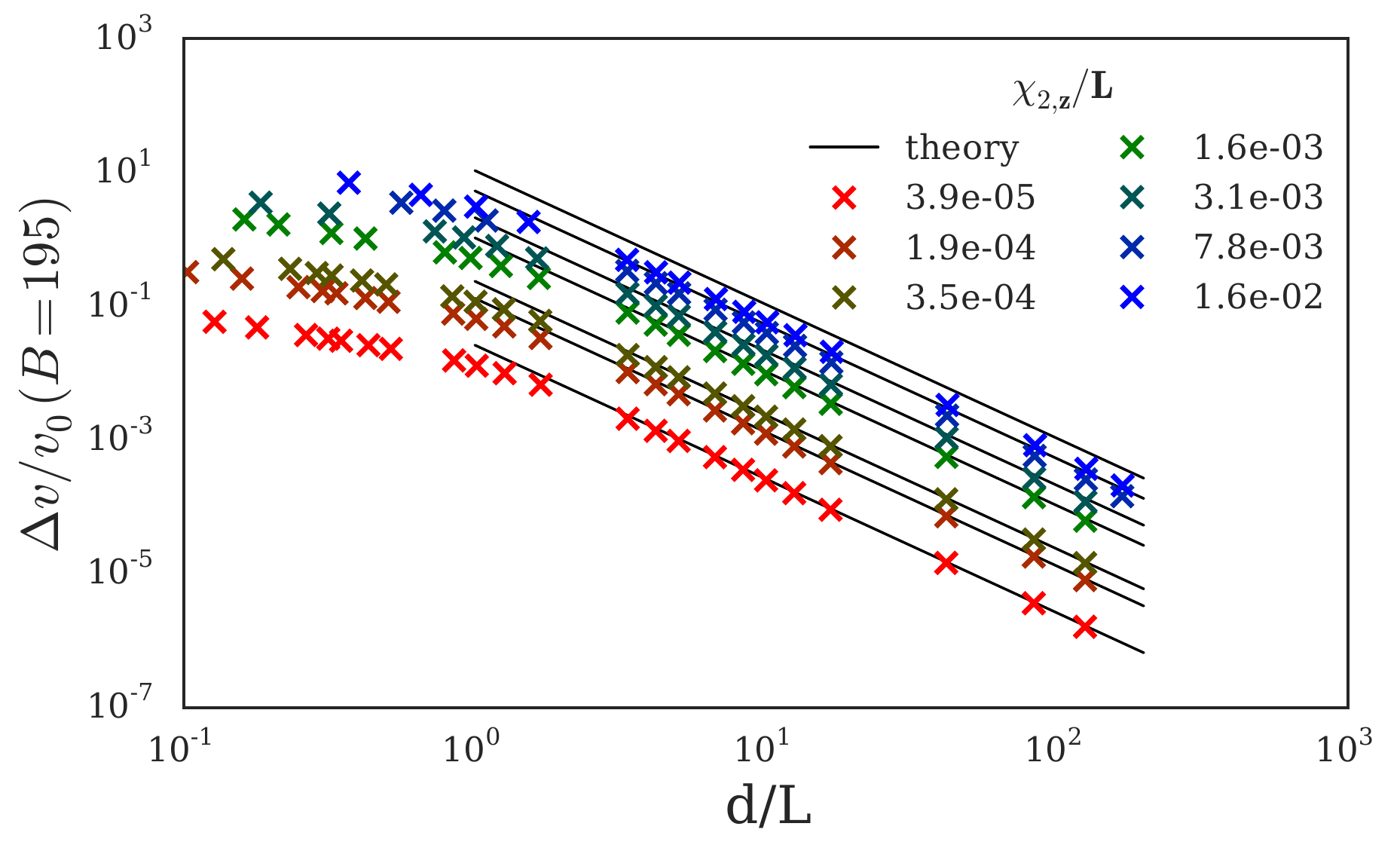}
{Simulations of two filaments  with the same imposed shape,  kept constant during the
simulation ($B=195$). The shapes are created with the given $\chi_{2,z}$. The filaments lie in
the same plane, parallel to the external field. The relative velocity $\Delta v$ scales
as $d^{-2}$. The black lines correspond to the prediction of Eq.~(\ref{eq:velocity}), save for a  common factor \GS{$\delta\approx 11$}. The theory describes correctly the trend
on $d$, and the trend on $\chi_{2,z}$ holds up to $\chi_{2,z}=8\times 10^{-3}$.}
{fixed_same_shape}

\subsubsection{Weak Field -- Stability}

\begin{figure*}
\centering
\includegraphics[width=1.5\columnwidth]{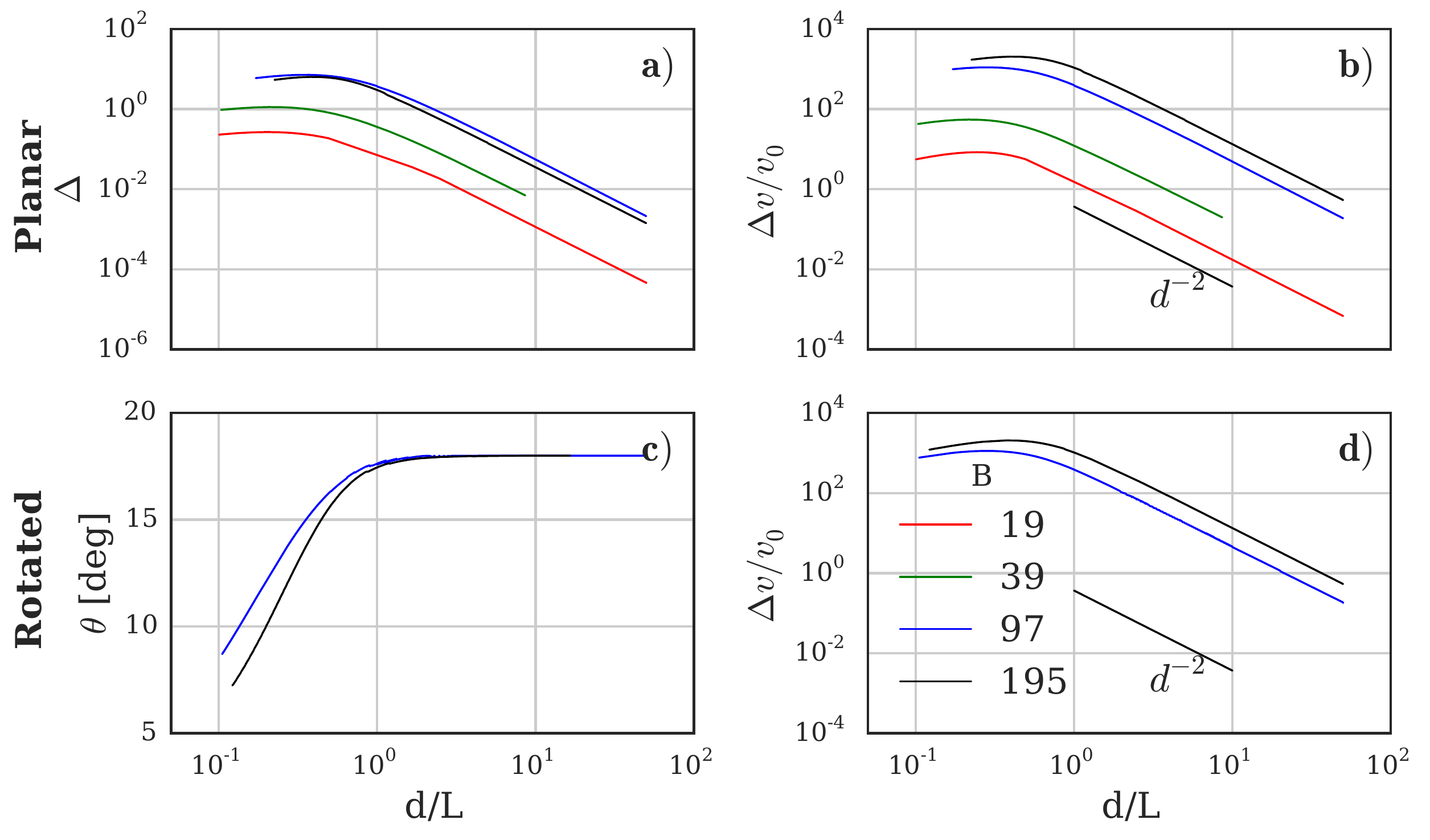}
\caption{
\textbf{a)-b)} Bending asymmetry $\Delta$ and relative
velocity $\Delta v$  of  two filaments as function of the filaments distance for $L/b=30$. The two filaments are in the same plane, parallel to the
external field and parallel to each other.  Each color corresponds to a different external field $B$, as indicated. The velocity
$v_0$ is the terminal velocity given by the resistive force theory for a rod. When $d/L\gg1$,
the relative velocity scales as $d^{-2}$. Note that filaments attract, i.e. time progresses
from right to left.
\textbf{c)-d)} Rotation angle $\theta$ and relative velocity $\Delta v$  of two  initially rotated filaments around the field axis
by $\theta=18^o$. The relative velocity is essentially unaffected by this change. Notably,
the filaments spin toward each other decreasing the relative angle.
\label{fig:res_geo2}
}
\end{figure*}

\coolimg{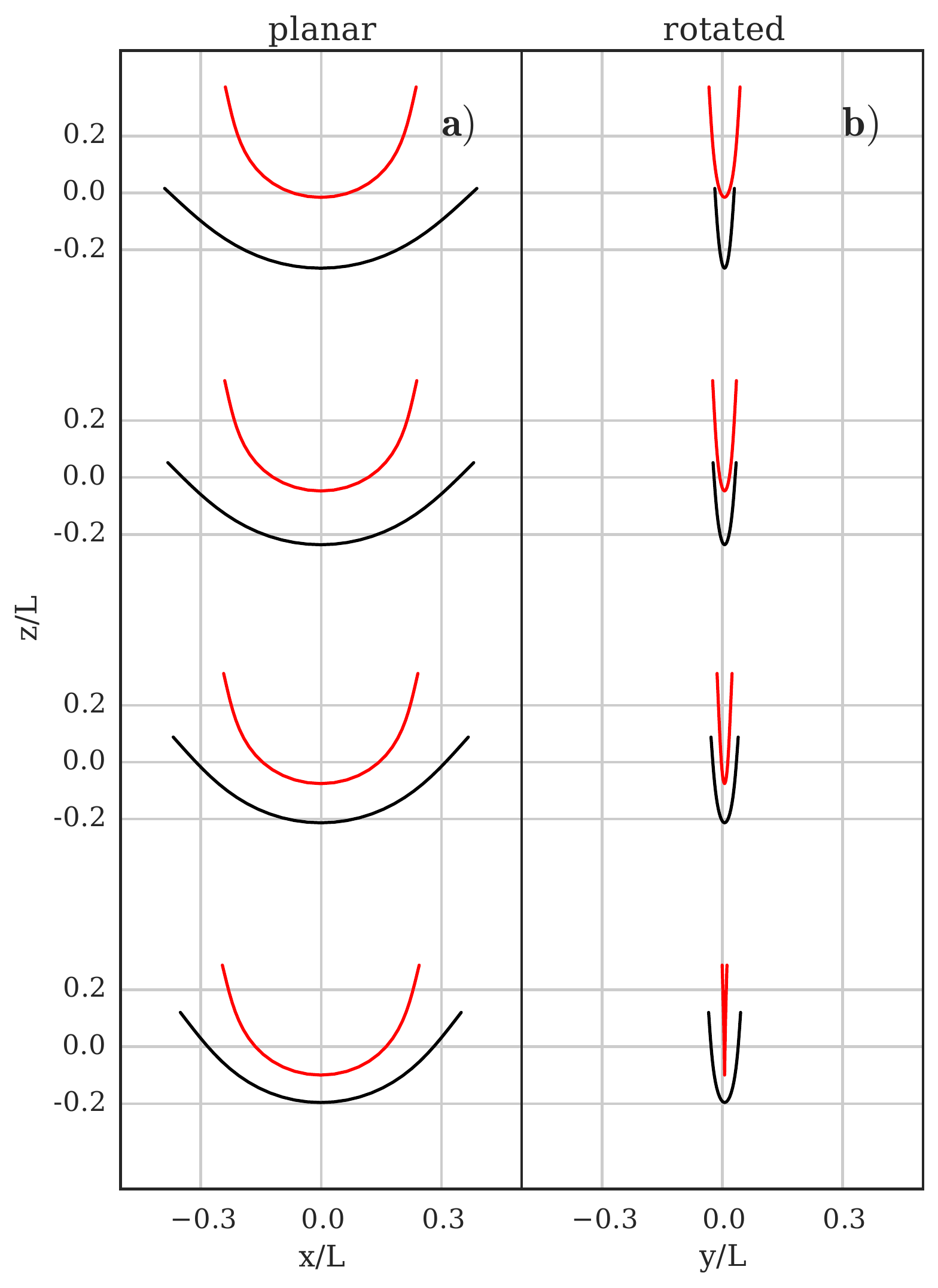}{Snapshots of two-filament conformations for $B=195$, in time intervals $\Delta t$. \textbf{(a)}
Co-planar sedimentation. Note that the upper filament is more bent than the lower filament, and
$d_{min}/L=0.13$. Axes to scale, $z$ position translated.
\textbf{(b)} The two
filaments approach each other after initialization in a rotated configuration. Both filaments
spin around the $z$ axis, with the upper filament spinning faster (see Fig.~\ref{fig:res_geo2}~(c-d)).}
{capture}

We now relax the imposed shape constraint and consider collective effects for two filaments, which are initially straight, oriented along the $x$ axis,  and displaced along the $z$ axis by a
distance $d$ (cf. Fig.~\ref{fig:capture}(a)).
For easier comparison with Ref.~\cite{cosentino_lagomarsino_hydrodynamic_2005,llopis_sedimentation_2007}, we  employ the dimensionless number $\Delta= (A_\text{upper}- A_\text{lower})/(L/2)$ to quantify
the bending asymmetry, where $A_\text{upper, lower}$ is the total $z$ extension of the
upper/lower filament. As indicated in Fig.~\ref{fig:res_geo2}, the filament curvature changes with time and the upper filament is bent stronger  than the lower one.   
Figures~\ref{fig:res_geo2} (a),(b), show the curvature asymmetries $\Delta$ and the relative velocities for various external field strengths. $\Delta$ decreases with increasing distance $d$, indicating more similar shapes at larger distances.
Hydrodynamic interactions lead to an attraction of the two filaments ($v_\text{upper}>v_\text{lower}$), in agreement with the imposed-shape approximation studies of the last section. Indeed, the constant-shape approximation still gives the
correct $(L/d)^{2}$ power-law  dependence for $d/L\gg 1$, while the magnitude of the deformation,
$\chi_{2,z}^{(eff)}$, has to be fitted. When the filaments approach each other, the generated flow field depends on the details of their shapes that, in turn, depends on the external field, hence we expect a
non-universal behavior. Note that in contrast to Ref.~\cite{llopis_sedimentation_2007}, we find
that the {\em upper} filament bends more than the lower filament (see also Fig.~\ref{fig:capture}).

The planar configuration of a filament is also stable with respect to filament rotations around the
field axis, see Fig.~\ref{fig:capture}~(b). Filaments that are initially displaced along the $z$ axis (as in the previous case) and rotated with relative orientation angle  $\theta$ around the external field axis spin  until the relative 
angle vanishes, as illustrated by Fig.~\ref{fig:res_geo2}~(c). Also in this case, the
{\em upper} filament drifts and rotates faster than the lower one, see Fig.~\ref{fig:res_geo2}~(d).
The relative velocity is essentially the same as in the planar case.

Thus, two filaments sedimenting in weak fields relax toward a stable planar configuration one
behind the other. The shape of the filaments is dominated by the second mode, pointing downwards, as shown in Fig.~\ref{fig:capture}.
This mode dominates and it breaks the mirror symmetry of the hydrodynamic interactions even for filaments of the same shape.
Note that, in contrast to the single filament case, the system does not reach a stationary state velocity or shape, since the upper filament is always faster than the lower filament until the filaments touch each other.

\subsubsection{Strong Field}

For strong fields, we consider two filaments, which are initially displaced by $6L$ along the
field direction.
We measure the shape eigenvalues when the distance is $5L$,
in the quasi-stationary regime, and find
that the eigenvalues exhibit the same behavior as those of a single filament. This means that for $B>B_1^*$ the dynamics of each filament is dominated by the local flow field and not by the interactions with the other filament.
Indeed, we find no correlations between the orientations of the out-of-plane components of the two filaments for $B>B_1^*$: the two filaments can by chance bend out-of-plane and drift in arbitrary directions.

When $B > B_2^*$, the filaments undergo the same transitions as a single filament: each
of them reaches the same stationary shape and rotation velocity as an isolated filament.
We find no correlations between the rotation directions of the two filaments:
some filaments spin in the same direction, others in opposite directions, with no preference.
This highlights the relevance of hydrodynamic interactions between two filaments for external fields
weaker than $B_1^*$. 
\GSdel{Stronger fields reduce the effects of hydrodynamic interactions}
\GS{At stronger fields, the non-planar configurations generate forces that compete with and dominate over hydrodynamic interactions among the filaments
and the emergent behavior is, essentially, the same as that of an isolated filament.}

\subsection{Three Filaments}

\coolimg{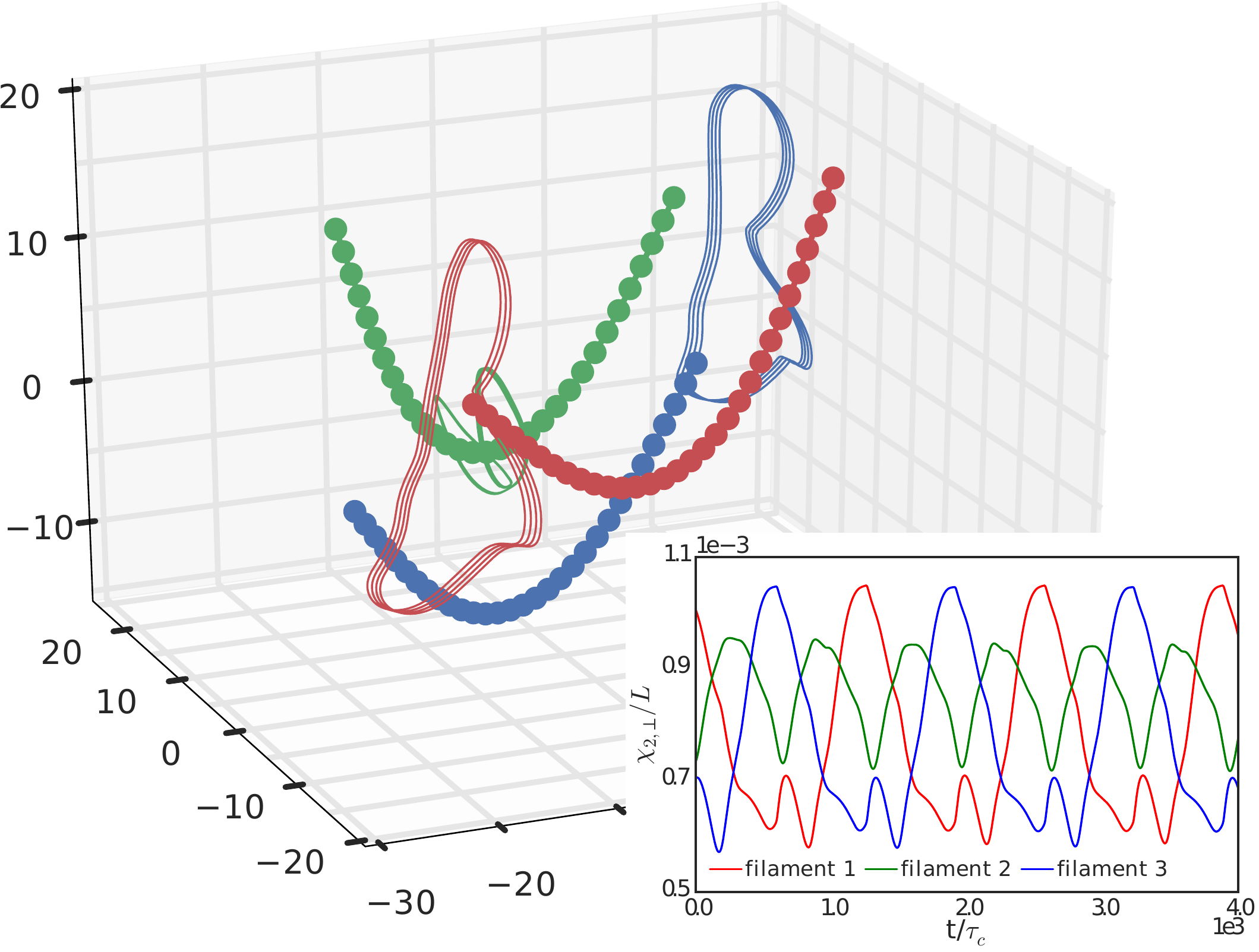}
{Three semiflexible filaments and trajectory of one bead (thick line), for the  external field
$B\simeq 60\ll B_1^*$. In this case,
the filaments form a bundle, but the relative positions  change periodically.
\textbf{Inset:} plot of $\chi_{2,z}$ for the three filaments. Since they have the same period
and constant phase shift,  this is the result of a cooperative behavior. Corresponding movie is shown in the ESI.}
{three}

Given the complex dynamics of two interacting filaments, it is interesting to consider also the
collective behavior of several filaments.
We find in simulations  of systems with more than two
filaments an intriguing collective dynamic behavior even for very weak fields ($B\simeq 60$) and in the absence of noise.

We focus here on the case of three filaments, see Fig.~\ref{fig:three}.
For most (randomly chosen) initial configurations, the nearest two filaments form a bundle
that settles faster than  the third filament that is then left behind.
%A similar behavior has been found for sedimenting colloidal particles \cite{ekiel-jezewska_stokesian_2008}.
However, we find also some initial configurations where all three
filaments attract each other and form a bundle. In this case, the relative
positions are not stationary; instead, the filaments follow a periodic trajectory, see
Fig.~\ref{fig:three} (inset).
In the inset of Fig.~\ref{fig:three}, we show also that the shapes of the three filaments are
not stationary. The mode amplitude $\chi_{2,z}$ of each filament changes periodically, with a
constant phase shift between them.

Our results for one and two filaments indicate that 
triggering of a time-periodic bifurcation requires strong fields. However, the three-filaments results suggest that systems with more filaments display a very complex dynamics even for weak fields due to complex hydrodynamic interactions.

\section{Discussion and Conclusions}

We have investigated the dynamics and stability of semiflexible filaments exposed to  an
external homogeneous field and interacting only via hydrodynamic fluid fields.  Due to the competition
between hydrodynamic interactions and bending stiffness, the appearing dynamical behavior is richer
than for entropy-dominated polymers or interacting rods.

We have shown that, for weak fields $B<B_1^*$, co-planar configurations of two filaments are stable
upon perturbations that rotate the shapes relative to each other around the field axis.
With simulations of fixed shape filaments, we have highlighted that a $V$- or $U$-shape is
sufficient to break the hydrodynamic symmetry at low Reynolds numbers, leading to a relative
velocity that scales with distance  as $(L/d)^{2}$. Hence, the difference in drag coefficients between filaments is not necessary to explain the faster settling velocity of the upper filament.

For external field strengths exceeding the critical value $B_1^*$, the hydrodynamic interactions bend the filament out of its principal plane.  Simulations of a single bent filament show that
the hydrodynamic forces balance the elastic force, stabilizing the out-of-plane shape.
The resulting trajectory shows a drift in the direction of out-of-plane bending, superimposed
to the settling motion.
This is a novel result, not to be confused with the previously reported metastable $W$-state
\cite{cosentino_lagomarsino_hydrodynamic_2005} that is excited when $B>B_2^*$.  A careful analysis of the eigenmodes  indicates that the decay of the metastable state does not, in general, lead to the reported planar horseshoe shape,
but also excites an average rotation mode with respect to the field axis ($\chi_{1,z}$) and
our out-of-plane bending mode $\chi_{2,\perp}$.
The filaments spin then around the field axis.

Finally, we have demonstrated that three filaments display an unexpected periodic dynamics even at field strengths far weaker than $B_1^*$. This is in contrast to
the dynamics of a pair of filaments that either displays a monotonic dynamics that relaxes the attractive force (when $B$ is weak) or a dynamics dominated by the single-filament (when $B>B_1^*$).

The interesting external fields $B$ are in the range  $10^1 \lesssim B \lesssim  10^4$. We
can estimate these parameters for  biopolymers like actin or microtubules.
Actin has a persistence length of $l_p\simeq 17 \mu m$, an average length $L\simeq20\mu m$, and the bending rigidity $\kappa\simeq 60\times 10^{-3} pN \mu m^2$.\cite{howard_mechanics_2001} The external gravitational field, corrected for buoyancy, is about $G \approx 10^{-7} \frac{\text{pN}}{\mu m}$, which implies $B_{gravity}\simeq 10^{-2}$.
Microtubules, on the other hand, are  stiffer, longer,  and heavier with $l_p\sim 1mm$, $L\sim100\mu m\ll l_p$ and $G\approx 10^{-6} \frac{\text{pN}}{\mu m}$.\cite{broedersz_modeling_2014}
This yields the effective field strength $B_{gravity}\simeq 10^{-1}$. An experimental test of our
predictions is therefore within reach of modern centrifuges with accelerations of about $10^3g$.

\appendix

\section{Eigenfunctions of a Semiflexible Filament} \label{app:eigenfunctions}

The eigenvalue equation (\ref{eq:eigenvalue}) with the boundary conditions
\begin{align}
\frac{\partial^2 }{\partial s^2} \phi_n(s)\left|_{s=\pm L/2} \right. = \frac{\partial^3 }{\partial s^3} \phi_n(s)\left|_{s=\pm L/2} = 0 \right.
\end{align}
yields the eigenfunctions 
\begin{align}
\phi_n(s) & = \frac{1}{\sqrt{L}} \left(\frac{\sinh \zeta_n s}{\sinh \zeta_n L/2} + \frac{\sin \zeta_n s}{\sin \zeta_n L/2} \right) , & n >1, \ \mathrm{odd} , \\
\phi_n(s) & = \frac{1}{\sqrt{L}} \left(\frac{\cosh \zeta_n s}{\cosh \zeta_n L/2} + \frac{\cos \zeta_n s}{\cos \zeta_n L/2} \right) , & n >1, \ \mathrm{even} .
\end{align}
The wave numbers are approximately given by $\zeta_n=(2n-1)\pi/2L$ ($n>1$), and the corresponding relaxation times
\begin{align}
\tau_n = \frac{16 \gamma L^4}{l_p \pi^4 k_B T (2n-1)^4} . 
\end{align}
More precise eigenfunctions are provided in Ref.~\cite{harnau_dynamic_1996,winkler_diffusion_2007}.
The set of functions is complemented by the eigenfunction of the center-of-mass translation \cite{winkler_diffusion_2007}
\begin{align}
\phi_0 = \frac{1}{\sqrt{L}}
\end{align}
and that of rotation of the rodlike object
\begin{align}
\phi_1= \sqrt{\frac{12}{L^3}} s ,
\end{align}
with the relaxation time
\begin{align}
\tau_1= \frac{\gamma L^3}{24 k_BT} .
\end{align}

\section{Relative Velocity of two Filaments} \label{app:relative_velocity}

We derive here an equation for the relative velocity between the centers of mass of two filaments. We restrict our analysis to the case of small bending amplitudes,
that is equivalent to consider small external fields, and filaments of identical shape.

Since  $\int_{-L/2}^{L/2}\phi_n(s)ds= \sqrt{L}\delta_{n,0}$ for the exact eigenfunctions, the difference in the center-of-mass velocity $\Delta \mathbf{v}_{cm} = \mathbf{v}_{cm}^1 - \mathbf{v}_{cm}^2$ of two isolated filaments is given by
\begin{align}
\Delta \mathbf{v}_{cm} =&\frac{1}{L}\int_{-L/2}^{L/2} ds\, \partial_t\big[ \mathbf r^1(s,t)-\mathbf r^2(s,t)\big] \nonumber\\
=& \frac{1}{\sqrt{L}} \partial_t \big[\bm\chi_{0}^1(t)-\bm\chi_{0}^2(t) \big].
\end{align}
Substitution of Eq.~(\ref{eq:modes_nonlinear}) yields
\begin{align*}
\sqrt{L} \Delta \mathbf{v}_{cm}=& 
\quad \sum_n \mathbf{H}_{0n}^{11} \left[-\frac{\gamma}{\tau_n} \bm\chi_{n}^1+ \mathbf{f}_{nG}^1 \right]  -  \sum_n \mathbf{H}_{0n}^{22} \left[-\frac{\gamma}{\tau_n} \bm\chi_{n}^2+ \mathbf{f}_{nG}^2 \right] \\
& + \sum_n \mathbf{H}_{0n}^{12} \left[-\frac{\gamma}{\tau_n} \bm\chi_{n}^2+ \mathbf{f}_{nG}^2 \right]  -\sum_n \mathbf{H}_{0n}^{21} \left[-\frac{\gamma}{\tau_n} \bm\chi_{n}^1+ \mathbf{f}_{nG}^1 \right] .
\end{align*}
The first two terms on the right-hand side account for self-interactions of the individual filaments, the other two terms for the hydrodynamic interactions between the
filaments.

We simplify our considerations by assuming identical shapes of the filaments, i.e., we set $\bm \chi_n^1 = \bm \chi_n^2 := \bm \chi_n$. Moreover, for the constant external force the relation applies $\mathbf{f}_{nG}^{\nu} =\mathbf{f}_{0G}^{\nu} \delta_{0n}$ independent of the particular filament. Hence, its contribution vanishes, which yields
\begin{align} 
\Delta \mathbf{v}_{cm}&= \frac{1}{\sqrt{L}} \sum_{n=1}^\infty \left( \mathbf{H}_{0n}^{21} - \mathbf{H}_{0n}^{12} \right) \frac{\gamma}{\tau_n } \bm\chi_n . \label{eq:dvm2}
\end{align}
We are primarily interested in the distance dependence of the relative center-of-mass velocity. Hence, we additionally neglect the dyadic term in the hydrodynamic tensor  (\ref{eq:Oseen}). Moreover, the local friction term vanishes in Eq.~(\ref{eq:dvm2}), and the hydrodynamic tensor can be written as
\begin{align}
\mathbf{H}_{0n}^{\nu \mu} = \frac{1}{8 \pi \eta } \int_{-L/2}^{L/2} \frac{\phi_n(s)\phi_0(s')}{|\mathbf{r}^{\nu}(s) - \mathbf{r}^{\mu}(s')|} ds ds' .
\end{align}
Using the eigenfunction expansion Eq.~(\ref{eq:eigenfunction_expan}), we obtain
\begin{align} \nonumber
\mathbf{r}^{\nu}(s) - \mathbf{r}^{\mu}(s') = & \Delta \mathbf{r}_{cm}^{\nu \mu} + \sum_{n=1}^{\infty} \bm \chi_n \left( \phi_n(s) - \phi_n (s')\right) \\
 = &  \Delta \mathbf{r}_{cm}^{\nu \mu} + \bm \Xi(s,s') .
\end{align}
With this definition, we obtain for $\Delta \mathbf{H}_{0n}^{12} = \mathbf{H}_{0n}^{21} - \mathbf{H}_{0n}^{12} $
\begin{align}
\Delta \mathbf{H}_{0n}^{12} = & \frac{1}{8 \pi \eta } \int_{-L/2}^{L/2} ds ds' \phi_n(s) \phi_0(s')\\ \nonumber
& \times  \left[
\frac{1}{|\Delta \mathbf{r}_{cm}^{21} - \bm \Xi(s,s')|} - \frac{1}{|\Delta \mathbf{r}_{cm}^{21} +\bm \Xi(s,s')|} \right]\,.
\end{align}
In the limit $d = |\Delta \mathbf{r}_{cm}^{21}| \gg |\bm \Xi(s,s')|$, Taylor expansion yields 
\begin{align}
\Delta \mathbf{H}_{0n}^{12} = & \frac{1}{4 \pi \eta } \int_{-L/2}^{L/2} ds ds' \phi_n(s)\frac{\bm \Xi(s,s') \cdot \Delta \mathbf{r}_{cm}^{21}}{d^3}  \phi_0(s'),
\end{align}
and hence, 
\begin{align}
\Delta \mathbf{v}_{cm} = & \frac{1}{4 \pi \eta \sqrt{L}} \frac{1}{d^2}   \sum_{n=1}^{\infty} \frac{\gamma}{\tau_n} \bm \chi_n \int_{-L/2}^{L/2} ds ds' \phi_n(s) \frac{\bm \Xi(s,s') \cdot \Delta \mathbf{r}_{cm}^{21}}{d}  \phi_0(s'). 
\end{align}
Substituting $x=s/L$ and setting $\gamma = 3 \pi \eta$ \cite{harnau_dynamic_1996}, yields
\begin{align}
\Delta \mathbf{v}_{cm} = & \frac{3}{4}\frac{L^2}{d^2} \sum_{n=1}^{\infty} \frac{1}{\tau_n} \frac{\bm \chi_n}{\sqrt L}\int_{-1/2}^{1/2}dx dx' \phi_n(x) \frac{\bm \Xi(x,x') \cdot \Delta \mathbf{r}_{cm}^{21}}{d} \phi_0(x') . 
\end{align}

Thus, the relative velocity decreases quadratically with the distance between the filaments. There is evidently no velocity difference when $\Delta \mathbf{r}_{cm}^{21}$ is perpendicular to $\bm \Xi(s,s')$. In particular, there is no force between two specifically aligned rods as long as their director $\bm \chi_1$ is perpendicular to $\Delta \mathbf{r}_{cm}^{21}$.

%%%REFERENCES%%%
\bibliography{newbib} 

\end{document}